\documentclass[%
 reprint,
superscriptaddress,
 amsmath,amssymb,
 aps,
prb,
]{revtex4-1}

\usepackage{amssymb}
\usepackage{fullpage}
\usepackage{graphicx}
\usepackage{bibunits}
\usepackage{wrapfig}
\usepackage{floatflt}
\usepackage{color}
\usepackage{xspace}
\usepackage{dsfont}
\usepackage[official,right]{eurosym}
\usepackage[top=0.8in, bottom=0.75in, left=0.75in, right=0.75in]{geometry}

\usepackage{graphicx}
\usepackage{dcolumn}
\usepackage{bm}
\usepackage{hyperref}

\usepackage{latexsym}
\usepackage{placeins}

\usepackage{multirow}

\newcommand{\KCUC} {K$_2$CuSO$_4$Cl$_2$\xspace}

\renewcommand{\Im}[1]{\mathrm{Im\{#1\}}}

\newcommand{\be}{\begin{equation} }
\newcommand{\ee}{\end{equation} }
\newcommand{\bea}{\begin{eqnarray} }
\newcommand{\eea}{\end{eqnarray} }

\newcommand{\mb}[1]{\mathbf{#1}}

\def\Xint#1{\mathchoice
   {\XXint\displaystyle\textstyle{#1}}%
   {\XXint\textstyle\scriptstyle{#1}}%
   {\XXint\scriptstyle\scriptscriptstyle{#1}}%
   {\XXint\scriptscriptstyle\scriptscriptstyle{#1}}%
   \!\int}
\def\XXint#1#2#3{{\setbox0=\hbox{$#1{#2#3}{\int}$}
     \vcenter{\hbox{$#2#3$}}\kern-.5\wd0}}

\def\dashint{\Xint-}

\begin{document}

\title{Finite-temperature correlations in a quantum spin chain near saturation}

\author{D. Blosser}
\email{dblosser@phys.ethz.ch}
\affiliation{Laboratory for Solid State Physics, ETH Z\"urich, 8093 Z\"urich, Switzerland}

\author{N. Kestin}
\affiliation{Department of Quantum Matter Physics, University of Geneva, 24 Quai Ernest Ansermet, 1211 Geneva, Switzerland}

\author{K. Yu. Povarov}
\affiliation{Laboratory for Solid State Physics, ETH Z\"urich, 8093 Z\"urich, Switzerland}

\author{R. Bewley}
\affiliation{ISIS Facility, Rutherford Appleton Laboratory, Chilton, Didcot, Oxon OX11 0QX, United Kingdom}

\author{E. Coira}
\affiliation{Department of Quantum Matter Physics, University of Geneva, 24 Quai Ernest Ansermet, 1211 Geneva, Switzerland}

\author{T. Giamarchi}
\affiliation{Department of Quantum Matter Physics, University of Geneva, 24 Quai Ernest Ansermet, 1211 Geneva, Switzerland}

\author{A. Zheludev}
\email{zhelud@ethz.ch}
\homepage{http://www.neutron.ethz.ch/}
\affiliation{Laboratory for Solid State Physics, ETH Z\"urich, 8093 Z\"urich, Switzerland}

\date{\today}

\begin{abstract}
Inelastic neutron-scattering and finite-temperature density matrix renormalization group (DMRG) calculations are used to investigate the spin excitation spectrum of the $S=1/2$ Heisenberg  spin chain compound K$_2$CuSO$_4$Cl$_2$ at several temperatures in a magnetic field near saturation. Critical correlations characteristic of the predicted $z=2$, $d=1$ quantum phase transition occurring at saturation are shown to be consistent with the observed neutron spectra. The data is well described with a scaling function computed using a free fermion description of the spins, valid close to saturation, and the corresponding scaling limits. One of the most prominent non-universal spectral features of the data is a novel thermally activated longitudinal mode that remains underdamped across most of the Brillouin zone.
\end{abstract}


\maketitle

\section{\label{sec:introduction}Introduction}

Quantum phase transitions in spin systems can often be viewed as a  condensation of bosonic quasi-particles with an external magnetic field playing the role of chemical potential and magnetization that of particle density. In three dimensions ($d=3$), such transitions are well understood in terms of magnetic Bose-Einstein condensation \cite{Giamarchi2008,Zapf2014}. Similar transitions with dynamical exponent $z=2$ can also occur in $d=1$, but possess some unique features. While it is impossible for true long range order to develop, correlation functions show a particularly striking zero scale-factor universality\cite{sachdev_senthil_shankar_qcp_z2}. Such physics can in principle be studied with cold atoms in optical lattices for which one-dimensional (1D) condensates of bosons have been obtained \cite{bloch_dalibard_zwerger_manybody_ultracold_gases,MoritzEsslinger2003,StofferleEsslinger2004, Kinoshita1125,ParedesBloch2004}. However, such systems pose problems of homogeneity, due to the confining potential, and of detailed control of the number of particles.
The simplest realization of such features is to be found in a Heisenberg spin chain near magnetic saturation. Such a quantum critical point (QCP) connects a fully polarized high-field state to a partially magnetized Tomogana-Luttinger spin liquid state at lower applied fields \cite{Giamarchi2004Book,Mourigal2013,Kohno2009,Lake2005,Haelg2015_CuDCl}.

In spin chains near saturation, recent studies addressed the universality of thermodynamic properties \cite{Kono2015,JeongRonnow2015,Rohrkamp2010,JeongRonnow2015} and quasistatic (low-frequency limit) correlation functions \cite{Kuhne2009,JeongRonnow2015,Kuhne2011}. Unfortunately, the much-anticipated universality of finite-temperature dynamic spin correlations remained elusive to date due to unique technical challenges. In addition to universal features, the excitation spectrum contains a wealth of non-universal components. To focus the spectroscopic measurements on a suitable window free of such non-universal contamination, one typically relies on comparisons with numerical simulations.

Particularly useful is the density matrix renormalization group (DMRG) \cite{white_dmrg_article1_breakthrough,white_dmrg_article2_breakthrough} (combined with time evolution \cite{vidal_tebd_precursor_pure_state_dynamics_breakthrough,daley_kollath_schollwock_tdmrg_pure_state_dynamics_breakthrough,vidal_tebd_pure_state_dynamics_breakthrough,white_feiguin_tdmrg_pure_state_breakthrougn}) which allows to compute the complete excitation spectrum with enough accuracy in strongly correlated low-dimensional systems at $T=0$ \cite{WardRuegg2017,BouillotGiamarchi2011,SchmidigerZheludevPRL2013,SchmidigerZheludevPRB2013}.
Such a success is mostly due to the low entanglement of the ground state \cite{vidal_tebd_precursor_pure_state_dynamics_breakthrough}. The problem is that spin chains near saturation additionally have strong and rather complex temperature-activated non-universal excitations.

During the past decade, algorithms for mixed state dynamics appeared under various names\cite{verstraete_garciaripoll_cirac_tdmrg_dissipative_breakthrough,vidal_zwolak_tebd_temperature,feiguin_white_Tdmrg_just_after}, pushing the entanglement growth problem to its limits. For some simple cases, the development of theses DMRG-based tools has allowed to extract equilibrium
correlations at finite temperatures (T-DMRG)\cite{barthel_tdmrg_optimized_schemes,schollwock_dmrg_long_notes}.
This provided a welcome guidance to experiments \cite{coira_giam_kollath_spin_lattice_relax_nmr_low_t,dupont_maxime_capponi_spin_lattice_relax_nmr_low_t,coira_giam_kollath_dimerized_2bePublished} especially in situations where temperature is too high for a brute force application of field theory.

In the present study we leverage this technique to interpret the results of high-resolution inelastic neutron scattering measurements of finite temperature correlations in the spin chain compound \KCUC \cite{Giacovazzo1976,Haelg2014}. Our approach allows us to overcome certain material-specific complications and to untangle the universal and non-universal components of the spectrum. The former show a temperature evolution that is consistent with expectations for the $z=2$,  $d=1$ QCP corresponding to
magnetic saturation. Close to this point quantum critical scaling is expected \cite{sachdev_senthil_shankar_qcp_z2} and the system goes to a free fermion fixed point \cite{chitra_giam_qcp_universal_function_ll}. This allows us to use a Fredholm determinant technique to compute the scaling function
exactly \cite{zvoranev_cheianov_giam_fredholm_string,korepin_slavnov_fredholm_string_scaling_function,cruz_goncalves_xy_1d_sxsx_correlations}.
In addition to this universal scaling part, the non-universal part includes a robust thermally-activated longitudinal mode, which remains visible and underdamped across the entire Brillouin zone and in an energy range far exceeding the temperature scale.

The plan of the paper is as follows. In Sec.~\ref{sec:experiment}, we present the studied compound and the experimental procedures and results. In Sec.~\ref{sec:DataAnalysisNumerics}, we present the effective model and the different approaches for studying the system of weakly coupled chains. Finally, in Sec.~\ref{sec:discussion}, we discuss the results and extract the universal scaling behavior. Some technical details can be found in the 
Appendixes.

\section{Experimental} \label{sec:experiment}

\subsection{Sample preparation and experimental setup} \label{sec:SamplePrepExpSetup}

We study the synthetic compound \KCUC \cite{Haelg2014}, also known as the natural mineral chlorothionite \cite{Giacovazzo1976}. It crystallizes in an orthorhombic $Pnma$ structure with room-temperature lattice constants $(a,b,c)=(7.73,6.08,16.29)$~\AA.  Antiferromagnetic spin chains run along the crystallographic ${a}$ direction. They are formed by Cu-Cl-Cl-Cu superexchange bridges between ${S=1/2}$-carrying Cu$^\mathrm{2+}$ ions with corresponding exchange constant ${J_\parallel /k_\mathrm{B}\sim 3}$~K\cite{Haelg2014}. Magnetic anisotropy was estimated to be two orders of magnitude weaker\cite{Haelg2014}, and in the present context is irrelevant. On the other hand, interchain interactions,  though weak, cannot be entirely neglected. In particular, they lead to long range antiferromagnetic ordering at ${T_N \lesssim150}$~mK \cite{Haelg2014}. By extrapolating the measured $(H-T)$ phase boundary for the ordered state, the $T\rightarrow 0$ saturation field is determined to be $H_0 = 4.47$~T.

The single-crystal sample used for the present study was grown as described in Ref.~\onlinecite{Haelg2014}. Inelastic neutron scattering experiments were performed on the time-of-flight (TOF) spectrometer \textsc{LET} at the ISIS facility \cite{Bewley2011}. The sample was mounted on an aluminum sample holder in a $^3$He-$^4$He dilution refrigerator with the ${(a,c)}$~crystallographic plane horizontal. A magnetic field of up to 9~T was applied along the ${b}$ direction, which we shall denote as the $z$ axis. The data was collected with a fixed incident neutron energy ${E_\mathrm{i}=2.2}$~meV. All the obtained data was analyzed using the \textsc{Horace} software package \cite{Horace}.

\KCUC has four equivalent Cu$^{2+}$ ions per unit cell, arranged equidistantly in the $(b,c)$ plane. For an assumed Heisenberg spin Hamiltonian, the spin-spin separation is thus halved along the $b$ and $c$ directions. Instead of the experimental momentum transfer $\mb{Q}$, we shall often use the corresponding notation for reduced wave vector transfer:
$\mathbf{q} = (q_\parallel,q_b,q_c) = \left(\mathbf{Q}\!\cdot\! \mathbf{a}, \,\frac{1}{2}\mathbf{Q}\!\cdot\! \mathbf{b}, \,\frac{1}{2}\mathbf{Q}\!\cdot\! \mathbf{c}\right)$.

\subsection{Measured quantities} \label{sec:MeasuredQuantities}

Neutron scattering probes the dynamic spin structure factor which is the Fourier transform of the spin-spin correlation function:
\begin{equation}
\label{eq:dynamic_struct_factor}
\mathcal{S}^{\alpha \beta}(\mathbf{Q},\omega) = \int_{-\infty}^{\infty} \! \mathrm{d}\mathbf{r}\,\mathrm{d}t \,  \mathrm{e}^{i(\omega t-\mathbf{Q} \cdot \mathbf{r})}   \langle {S}^\alpha(\mathbf{r},t)   {S}^\beta(0,0)  \rangle,
\end{equation}
where $\langle \cdots \rangle$ denotes both the quantum and thermal average.
The obtained experimental data is processed as described in Appendix \ref{sec:unpolarized_neutron} such that the corrected scattering intensity is proportional to the following weighted sum of spin structure factors:
\begin{equation}
\mathcal{I} \propto  \frac{1}{2}\left(  \mathcal{S}^{xx}    +    \mathcal{S}^{yy}      \right)     +  \frac{{Q^2}-{Q_z^2}}{{Q^2}+{Q_z^2}} \,\mathcal{S}^{zz}.
\end{equation}
Note that this quantity is directly proportional to the transverse spin structure factor and only the weight of the longitudinal contribution is dependent on momentum transfer. However, the experimental coverage of momentum transfer along the $b$ axis is severely restricted due to the construction of the cryomagnet and to a very good approximation 
\begin{equation}
\mathcal{I} \propto \frac{1}{2}\left(  \mathcal{S}^{xx}    +    \mathcal{S}^{yy}      \right)     +  \mathcal{S}^{zz}.\label{eq:SumOfStructureFactorComponents}
\end{equation}

\subsection{Results} \label{sec:Results}

\begin{figure}
\includegraphics{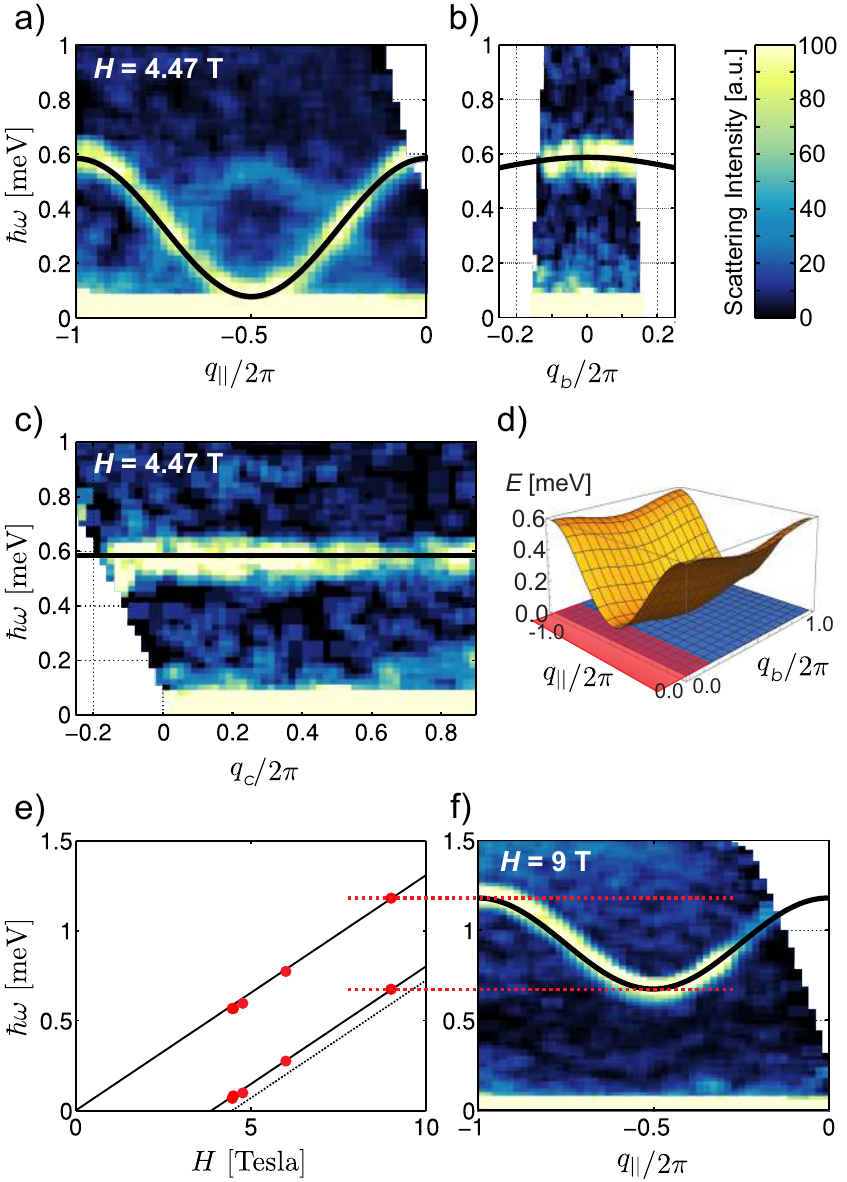}
\caption{\label{fig:GapAndDispersion} Magnon dispersion measured in \KCUC at ${T=0.2}$~K: a-c) False-color intensity plots along the principal reciprocal space directions at saturation ($H_0=4.47$~T). In a,b) the intensity is integrated in the range ${q_b/ 2\pi = 0 \pm 0.05}$ or ${q_\parallel/2\pi=-1\pm 0.05}$, respectively, and completely along the non-dispersive direction $q_c$. For c) the integration slice is ${q_\parallel/ 2\pi =-1\pm 0.1}$ and ${q_b/2\pi = 0\pm 0.1}$.
Solid lines indicate the dispersion calculated for the minimal model of the spin Hamiltonian described in the text. The full 2D dispersion of this minimal model at saturation is shown in d), where the approximate volume of reciprocal space covered in our measurements is indicated in red. e) Measured field dependence of the dispersion minima and maxima at $q_b=0$ (symbols). Solid lines correspond to the minimal model. The dashed line corresponds to the true minimum of the 2D dispersion at $q_\parallel = \pi$, $q_b=\pi$. f) As in a), for $H=9$~T.}
\end{figure}

\begin{figure*}
\includegraphics{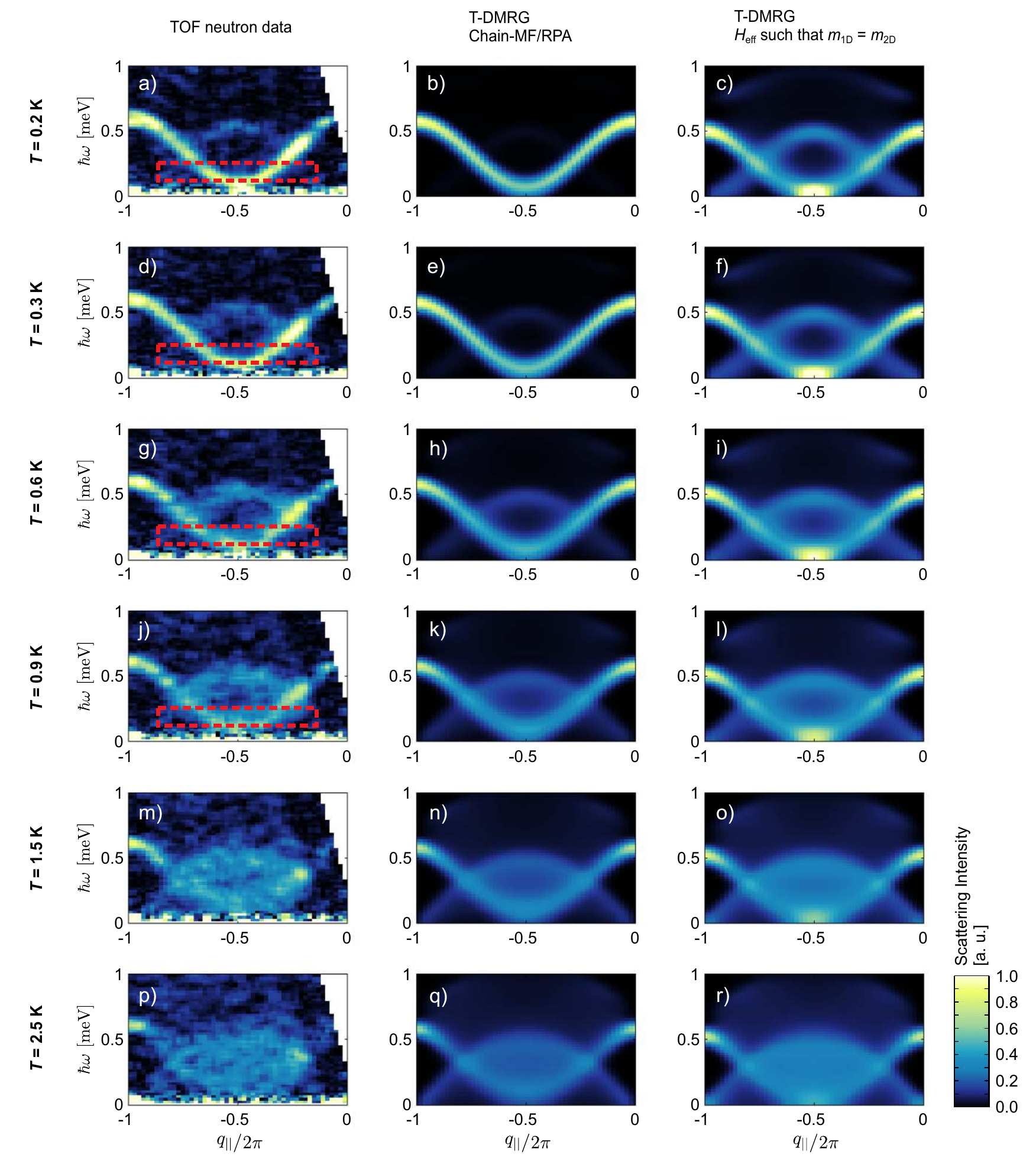}
\caption{\label{fig:Spectra} Measured and calculated spin excitation spectra of the Heisenberg spin chain compound \KCUC near saturation. The first column shows the inelastic neutron scattering data collected very close to saturation at $H_\mathrm{exp} = 4.5$~T  ($H_0=4.47$~T) at different temperatures. These plots correspond to slices integrated in the range $q_b/ 2\pi = 0 \pm 0.1$ and completely along the non-dispersive direction $q_c$. The red dashed rectangles delineate the regions of the spectra that are analyzed in regard to universal behavior in Sec.~\ref{subsubsec:Scaling}. The second column shows numerical results where interchain exchange is treated within a combined chain-MF/RPA approach. In the last column the results of a purely one-dimensional calculation are plotted. For this calculation the effective magnetic field was chosen such that the magnetization of the simulated 1D chains exactly corresponds to the magnetization of the two-dimensional system of weakly coupled chains at the experimentally applied magnetic field.}
\end{figure*}

\begin{figure*}
\includegraphics{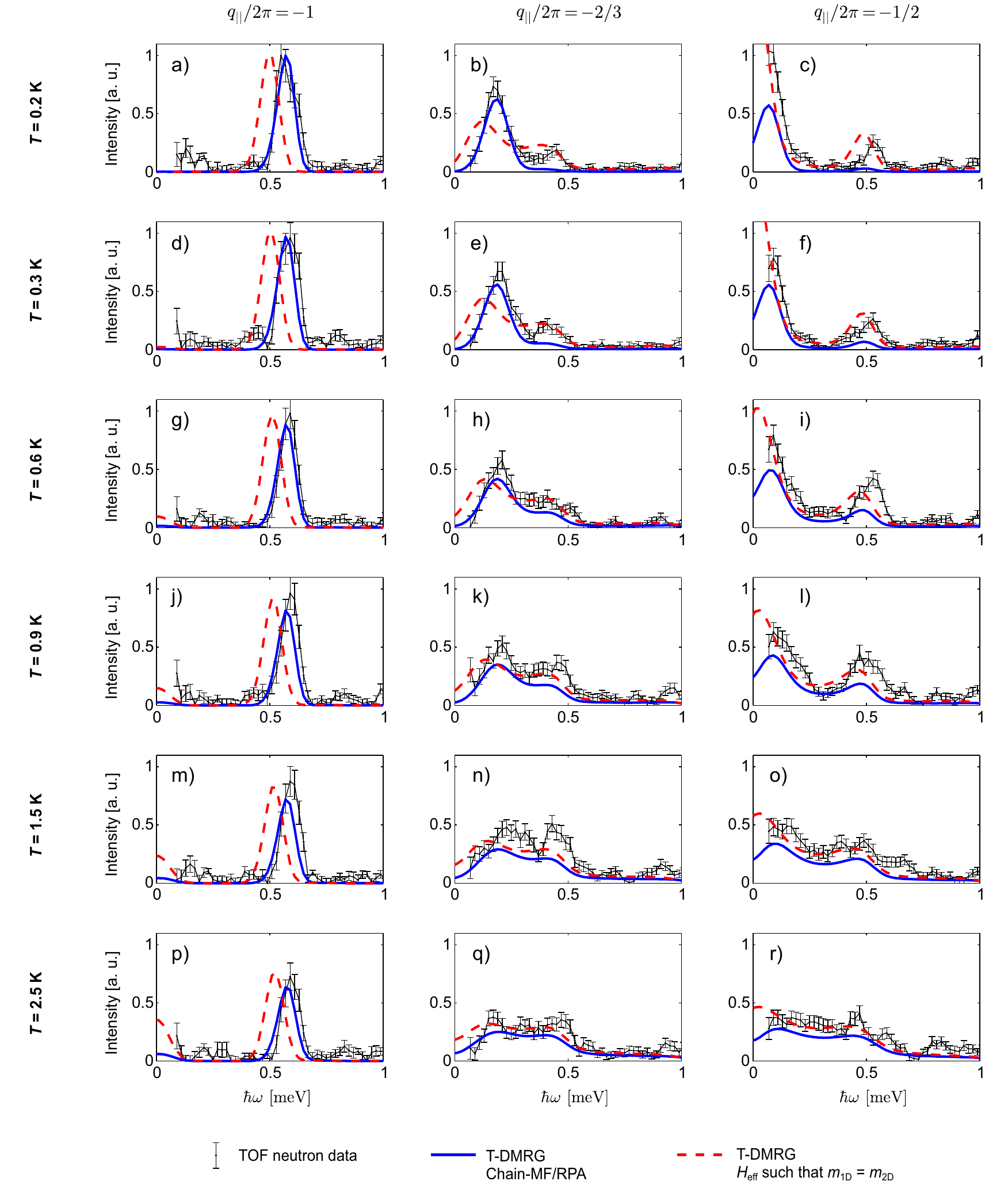}
\caption{\label{fig:Cuts} Cuts through the data shown in Fig.~\ref{fig:Spectra} at momenta $q_{\parallel}/2 \pi=\{-1,-2/3,-1/2\} $. The neutron data is plotted only for energy transfers $\hbar \omega >0.06$~meV ($\hbar \omega >0.08$~meV at $q_\parallel/2 \pi=-1$) so as to avoid contributions due to incoherent scattering (tails of the ($-$1,0,0) Bragg peak). Error bars for the neutron data correspond to $\pm1$ standard deviation. }
\end{figure*}

The bulk of our neutron scattering data can be grouped in two sets. Those in the first set were taken at base temperature $T=0.2$~K and are shown in Fig.~\ref{fig:GapAndDispersion}. The non-magnetic background is 10--20 times lower than the magnetic signal and no background was subtracted. These data were primarily used to establish the exact spin Hamiltonian as described in Sec.~\ref{sec:SpinHamiltonian}. Data in the second set were collected at ${H_\mathrm{exp}=4.5}$~T for six different temperatures in the range ${T=0.2-2.5}$~K, to study the thermal evolution of the spectrum near saturation. They are presented as false color plots in the first column of Fig.~\ref{fig:Spectra} and as cut-out scans in Fig.~\ref{fig:Cuts}. For these data sets, the non-magnetic background was obtained from base-temperature measurements at ${H=9}$~T and subtracted point by point. The normalization is arbitrary but the same at all temperatures. It is such that the maximum peak scattering intensity at ${q_{\parallel}/2\pi=-1}$ at the lowest temperature {($T=0.2$~K)} is unity.

At low temperatures the most prominent feature in the measured spectra is a well-defined magnon with a $\cos$ dispersion along the chains and a minimum at ${q_{\parallel}=\pi}$. Somewhat unexpected is a second, well-defined but much weaker mode, whose dispersion at $q_{\parallel}=\pi$ shows a maximum. At a first glance, it seems to be a shifted replica of the magnon excitation that could result from Brillouin zone folding. Such folding occurs in some other quasi-1D materials such as BaCu$_2$Si$_2$O$_7$ \cite{ZheludevKenzelmann2001} due to zigzag spin chains.  However, a more careful examination of the \KCUC data reveals that the dispersion of the two excitations does not match. Moreover, even though there are four Cu$^{2+}$ ions in the unit cell of \KCUC, they are all in high symmetry position. There is no zigzagging of the spin chains, and no folding of the Brillouin zone as far as magnetic excitations are concerned. As will be discussed below, this additional mode has an entirely different origin and is a unique \emph{longitudinal} excitation characteristic of a Heisenberg ${S=1/2}$ chain near saturation. From our experiments, we also see that at a finite temperature both the magnon and the longitudinal mode broaden in energy, but in an asymmetric fashion. The former progressively weakens with increasing $T$ while the latter appears to strengthen. As a result, at high temperatures the scattering can be described as a lens-shaped continuum.

\section{Data Analysis and Numerical calculations} \label{sec:DataAnalysisNumerics}

\subsection{Spin Hamiltonian} \label{sec:SpinHamiltonian}
To understand this rich experimentally observed behavior we first need to establish the effective spin Hamiltonian for \KCUC.
Since at $T\rightarrow 0$ in the fully saturated phase spin wave theory becomes exact, the magnon dispersion relation directly provides the actual un-renormalized exchange constants \cite{Coldea2002}. For \KCUC we performed this analysis using spectra collected at base temperature at and above saturation, as shown in Fig.~\ref{fig:GapAndDispersion}.
As discussed in Ref.~\onlinecite{Haelg2014}, the magnon dispersion along the  $a$ (chain) direction can be fully accounted for by nearest-neighbor Heisenberg interactions with an exchange constant $J_{\parallel}$.
From our present data we obtain ${J_{\parallel}/k_\mathrm{B}=2.94\,\mathrm{K}}$, fully in agreement with previous estimates. However, our measurements reveal an additional feature of the magnon dispersion. Even though ${H_0=4.47}$~T exactly corresponds to the saturation field, in Fig.~\ref{fig:GapAndDispersion}a there still appears to be a small excitation gap $\Delta$ at $q_\parallel=\pi$, $q_b=q_c=0$.
This apparent gap can be accurately determined by a linear fit (Fig.~\ref{fig:GapAndDispersion}e) of the Zeeman-shifted magnon energies measured at several higher fields (Fig.~\ref{fig:GapAndDispersion}f).
The result of such a linear fit at $H_0$ is $\Delta=78$~$\mathrm{\mu}$eV. 
The most obvious explanation for the observed excitation gap would be, that the applied magnetic field is actually too high and the saturation field is $H=3.87$~T where the observed magnon gap extrapolates to zero. However, data obtained in a short measurement at $H=4.07$~T (presented in appendix \ref{sec:BelowSaturation}) clearly corresponds to a partially magnetized chain \cite{Kohno2009} and this possibility can be excluded.
Dzialoshinzkyi-Moria (DM) interactions have been estimated to be very small\cite{Haelg2014}: $|\mathbf{D}|\approx 0.014 J$ and within the present experimental energy resolution such effects are certainly not observable.

Thus, the spectrum at $H_0$ must be gapless, and our apparent gap simply means that the actual minimum of the three dimensional dispersion relation is not at $q_\parallel=\pi$, $q_b=q_c=0$. Therefore, interchain coupling must be stronger than anticipated. In our experiments, there is clearly no detectable dispersion along $q_c$ (Fig.~\ref{fig:GapAndDispersion}c). The data provide an upper bound on any exchange interactions in that direction: $J_c/k_\mathrm{B} < 0.1\,\mathrm{K}$: quite negligible for the following discussion. Unfortunately, a direct assessment of $J_b$ is prevented by a restrictive scattering geometry due to the construction of the cryomagnet. If we assume a single AF exchange constant $J_b>0$, at saturation the gap is closed at $q_\parallel=\pi$, $q_b=\pi$. At the same time, the excitation energy at $q_\parallel=\pi$, $q_b=0$ will be $\Delta=2 J_b$. This interpretation for the observed gap yields $J_b/k_\mathrm{B}=0.45\,\mathrm{K}$. It also explains why the saturation field in \KCUC is larger than that for an isolated spin chain with the same exchange constant and $g$ factor,  $H_c^\mathrm{1d}=\frac{2 J_{\parallel}}{g \mu_\mathrm{B}}=3.87$~T. Somewhat unexpectedly, the appropriate model for \KCUC is thus a quasi-two-dimensional system composed of weakly coupled spin chains:
\begin{align} \label{eq:Hamiltonian}
\mathbf{H}_{2d} =& J_{\parallel} \sum_{i,n} \mathbf{S}_{i,n} \cdot \mathbf{S}_{i+1,n}
+  J_b \sum_{i,n} \mathbf{S}_{i,n} \cdot \mathbf{S}_{i,n+1} \notag \\
&- h^z_\mathrm{exp} \sum_{i,n}S^z_{i,n}.
\end{align}
where $h^z_\mathrm{exp} = g \mu_\mathrm{B} \mu_0 H_\mathrm{exp}$. The index $n$ labels the
chains and $i$ the spins in each chain, respectively. The parameters read
\begin{equation}
 J_{\parallel}/k_\mathrm{B}=2.94\,\mathrm{K}, \quad  J_{b}/k_\mathrm{B}=0.45\,\mathrm{K}, \quad g=2.26.
\end{equation}

Note that our value for $J_b$ is an order of magnitude larger than the one quoted in Ref.~\onlinecite{Haelg2014}. The latter was estimated indirectly, from the measured $T_N$ and an application of chain-MF theory, and not from any spectroscopic data.
That analysis becomes intrinsically flawed in the quasi-two-dimensional case, where long range order is suppressed. The previous $T_N$-based estimate is therefore superseded by our present more direct measurement. Moreover, the magnon dispersion relation calculated with our present set of parameters (solid lines in Figs.~\ref{fig:GapAndDispersion}a,b,c,f) is consistent with the gently downward-curving shape of the scattering in the $(\hbar \omega,q_b)$ plane (Fig.~\ref{fig:GapAndDispersion}b).

\subsection{\label{sec:1d_frame}Effective 1D Spin Chain}

The presence of the (not so) weak interchain exchange $J_b$ makes it more challenging to isolate universal one-dimensional features and to identify the nonuniversal ones in our experimental data. Guidance from numerical calculations thus becomes critical.
Note that those calculations also need to somehow include interchain coupling. Given the fact that we are ultimately interested
in real-time dynamics, standard numerical methods such as quantum Monte Carlo, which could be used directly in two dimensions but which compute the dynamics in imaginary time are problematic because of the need of analytic continuation.

In order to use DMRG that gives access to the real-time dynamics we thus first need to transform the problem into an effective
1D Heisenberg spin chain
\begin{equation}
  \label{eq:1dheisenberg}
\mathbf{H}_\mathrm{1d} = J_{\parallel} \sum_{i} \mathbf{S}_{i} \cdot \mathbf{S}_{i+1}- h^z \sum_{i}S^z_{i}.
\end{equation}
In order to do so we employ the two following techniques.

\subsubsection{Chain mean field and random phase approximation} \label{subsec:methods_rpa}

Our first approach to dealing with interchain coupling is the well established\cite{Schulz1996,EsslerTsvelikDelfino1997,Zheludev2003,BouillotGiamarchi2011} chain Mean Field (Chain-MF) and Random Phase Approximation (RPA). The interchain interactions are treated by mean field and as a direct consequence, the effective magnetic field is given by
\begin{equation} \label{eq:meanfieldmag}
 h^z_\mathrm{MF} = h^z_\mathrm{exp} - J_b \mathcal{Z} m,
\end{equation}
where $\mathcal{Z}=2$ is the number of neighboring chains and $m$ the magnetization per site. This magnetization can be taken from the experiments themselves or from the solution of the full two-dimensional problem. Here, it is computed self-consistently
using the complete 1D magnetization curve as a function of the external magnetic field and temperature.
\begin{equation} \label{eq:selfcons}
m_{\mathrm{1D}}\left(T, h^z_\mathrm{MF} \right) = m.
\end{equation}
The six magnetization curves $m_{\mathrm{1D}}(T, h^z)$ were numerically computed from the T-DMRG simulations (see Appendix~\ref{sec:processing}) for the model (\ref{eq:1dheisenberg}).
In Table~\ref{tab:MagnetizationValues} we list the effective fields $h^z_\mathrm{MF}$ computed for the conditions of our experiments at the actual applied field $H_\mathrm{exp}=4.5$~T and various temperatures.

\begin{table}
  \centering
  \begin{tabular}{ l c c c l l c c c l   l c  c  c }
  \hline\hline
   & \multicolumn{3}{c}{Experiment}         &   &   &\multicolumn{3}{c}{Chain-MF/RPA}     &   &     & \multicolumn{3}{c}{ \begin{tabular}{@{}c@{}}Calculation with\\$m_{\text{2D}}=m_{\text{1D}}$\end{tabular}}\\
  \cline{2-4}\cline{7-9}\cline{12-14} \\[-8pt]
   & $h^{z}_{\text{exp}}$ [K] &   & $T$ [K] & \hspace{0.01\textwidth}  &   & $m$     &   & $h^{z}_{\text{MF}}$ [K] &  \hspace{0.01\textwidth}  &     & $m$      &    & $h^{z}_{\text{eff}}$ [K]\\
   \\[-8pt]
   \cline{2-4}\cline{7-9}\cline{12-14}    \\[-9pt]
   & 6.83                     &   & 0.2     &   &   & 0.493 &   & 6.39                    &   &     & 0.429  &    & 5.81                    \\
   & 6.83                     &   & 0.3     &   &   & 0.481 &   & 6.40                    &   &     & 0.426  &    & 5.86                    \\
   & 6.83                     &   & 0.6     &   &   & 0.446 &   & 6.43                    &   &     & 0.405  &    & 5.94                    \\
   & 6.83                     &   & 0.9     &   &   & 0.418 &   & 6.46                    &   &     & 0.384  &    & 5.98                    \\
   & 6.83                     &   & 1.5     &   &   & 0.373 &   & 6.50                    &   &     & 0.348  &    & 6.04                    \\
   & 6.83                     &   & 2.5     &   &   & 0.319 &   & 6.55                    &   &     & 0.298  &    & 6.08                    \\
  \hline\hline
\end{tabular}
\caption{\label{tab:MagnetizationValues}
Values of the uniform magnetization for various temperatures $T$ and the experimental magnetic field $h^z_\mathrm{exp}$, as shown in the left column. The central column gives the value of the magnetization $m$ and the effective magnetic field $h^z_\mathrm{MF}$ determined by a self consistent solution of the mean-field one dimensional model (\ref{eq:selfcons}). The right column gives the effective magnetic field $h^z_\mathrm{eff}$ determined from  (\ref{eq:m1m2}) by using the magnetization $m_\mathrm{2D}$ of the full two-dimensional system. These two procedures quite naturally lead to different values of the effective magnetic field (which should be compared to ${h_c^\mathrm{1D}=2 J_\parallel =5.88\,\mathrm{K}}$). Note that in this table all values of magnetic field are reported in
units of Kelvin to allow direct comparison.
}
\end{table}

\subsubsection{Calculations at correct magnetization } \label{subsec:methods_m1m2}

The chain-MF approximation is known to have a number of intrinsic limitations. Its greatest flaw is that by ignoring 2D correlations, it systematically overestimates the magnetization. This is a problem because the latter relates to the boson density, which is a key critical quantity of the QCP. A second approach aims to overcome this deficiency. We start with the true magnetization $m_\mathrm{2D}(T,h^z_\mathrm{exp})$ of the complete two-dimensional model of weakly coupled chains (\ref{eq:Hamiltonian}). We then use T-DMRG simulations to find an effective field $h^z_\mathrm{eff}$, which at the given temperature would induce exactly the correct magnetization value in a single spin chain:
\begin{equation}\label{eq:m1m2}
  m_{\mathrm{1D}}( T, h^z_\mathrm{eff})=m_\mathrm{2D}(T,h^z_\mathrm{exp}).
\end{equation}
We list in Table~\ref{tab:MagnetizationValues} the effective magnetic fields $h^z_\mathrm{eff}$ obtained with such a procedure.

Note that there are small differences between the two procedures as can be anticipated. These differences are washed out in the high-temperature regime.

\subsubsection{Calculations of the structure factor and comparison with experiments} \label{subsec:structurecomparison}

\begin{figure*}
\includegraphics{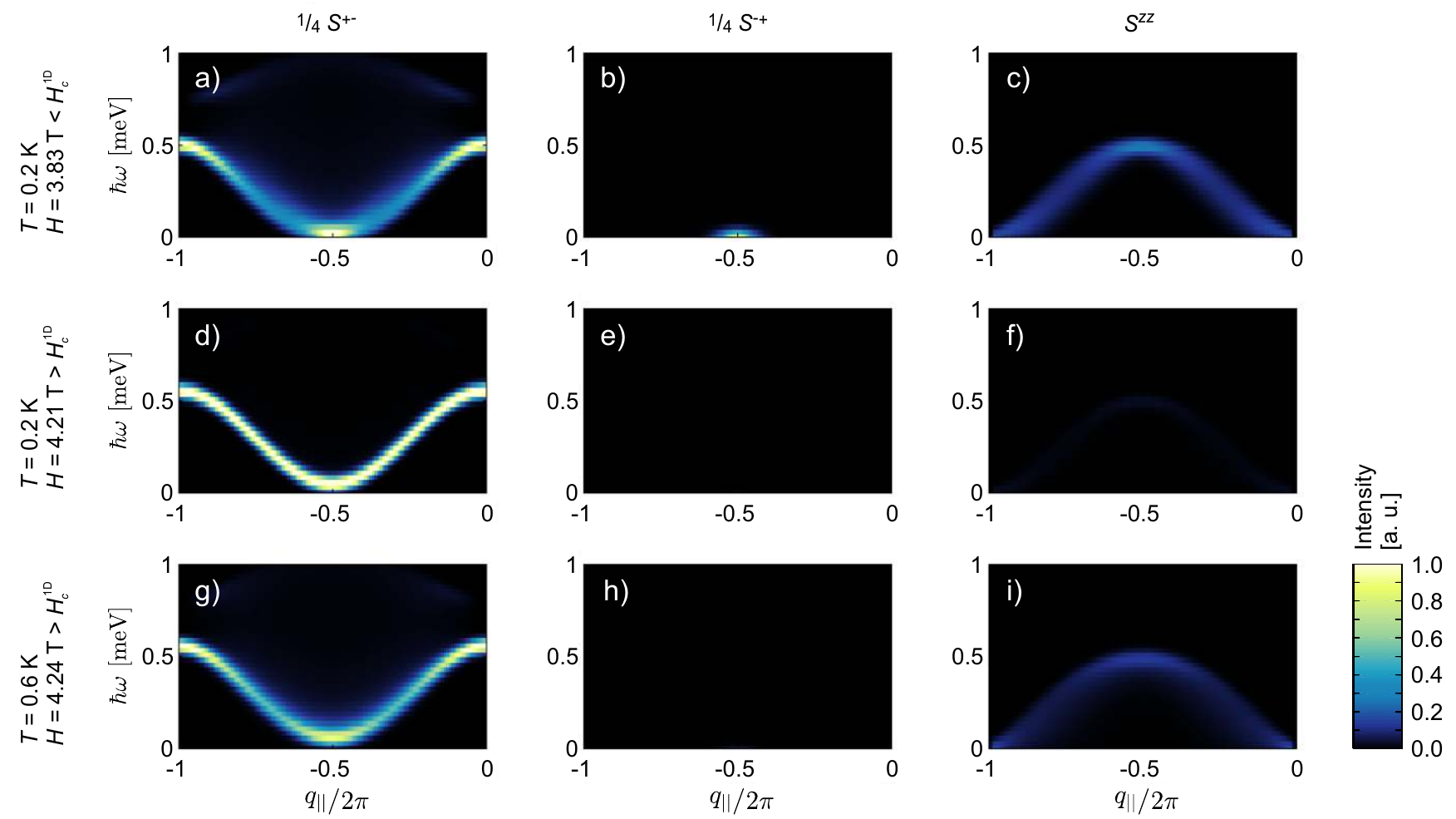}
\caption{\label{fig:polarization}Calculated single-chain excitation spectra with the different polarization channels plotted separately. a-c) Low temperature slightly below saturation, d-f) low temperature above saturation and g-i) elevated temperature above saturation. While in the fully polarized regime at ${H>H_c^\mathrm{1D}=\frac{2J_\parallel}{g \mu_\mathrm{B}}=3.87}$~T no longitudinal excitations are possible at zero temperature, at elevated temperatures larger than the magnon gap a thermally activated longitudinal mode gains spectral weight.}
\end{figure*}

With the one dimensional effective magnetic field fixed we can use the T-DMRG on the 1d model (\ref{eq:1dheisenberg}) to compute the entire dynamic susceptibility $\chi^{\alpha\beta}_\mathrm{1D}(q,\omega)$ of the single chain as described in Appendix~\ref{sec:chi}. Since the mean-field approximation only takes care of the static effects of the transverse coupling we also use the RPA for this case to give a better description of the full response:
\begin{equation}
\chi^{\alpha \beta}_\mathrm{2D}(\mathbf{q},\omega) = \frac{\chi^{\alpha \beta}_\mathrm{1D}(q_{\parallel},\omega)}{1 - 2 \, J_b \cos(q_b) \, \chi^{\alpha \beta}_\mathrm{1D}(q_{||},\omega)}.
\end{equation}
for $\alpha\beta \in \{xx,yy,zz\}$. From this, the dynamic structure factor (\ref{eq:dynamic_struct_factor}) measured by the neutrons is then obtained as explained in Appendix~\ref{sec:chi}.

While in the experimental data the different polarizations cannot be separated, numerically they are calculated separately. In Fig.~\ref{fig:polarization}, we separately plot the different spectral components of calculated single-chain excitation spectra.

To enable a direct comparison with experiment, the numerical results were also convoluted with the known experimental energy and momentum resolution. For all simulations the same overall normalization is used as for the experimental data. The corresponding results are shown in Fig.~\ref{fig:Spectra} for the two ways to determine the effective magnetic field discussed in Table~\ref{tab:MagnetizationValues}.
Cuts comparing the data (black) and the theoretical curves (respectively, the solid blue and dashed red lines) are shown in Fig.~\ref{fig:Cuts}.

\section{Discussion}\label{sec:discussion}

\subsection{Components of the excitation spectrum}\label{subsec:SpectralComponents}

Using the numerical calculations we are now in a position to analyze the experiment. In particular, and contrarily to the experimental data for which all components are present for unpolarized neutrons, we can separate the contributions of the various components in the numerics as shown in Fig.~\ref{fig:polarization}.

As expected, the spin excitation is transversely polarized at low energy. At the same time, it becomes clear that the additional mode observed in the experiment is a longitudinal one. We stress that this feature is present in numerical simulations of the idealized one-dimensional model, and is not a result of interchain coupling or zone folding.

The simulation shown in Fig.~\ref{fig:polarization}a,b,c) corresponds to Fig.~\ref{fig:Spectra}c. It was performed at  $H=3.83$~T, i.e., below the saturation field of a single chain $H_c^\mathrm{1D}=3.87$~T. For this regime just below saturation, a narrow dispersive stripe of longitudinal excitations has indeed been predicted \cite{Kohno2009}, in full agreement with our calculations. In the fully saturated state of the Heisenberg model though, no longitudinal excitations can exist. This is indeed what we observe in a simulation at $H=4.21\;\mathrm{T}>H_c^\mathrm{1D}$ at very low temperature as shown in Fig.~\ref{fig:polarization}d,e,f).

However the single-chain simulation in Fig.~\ref{fig:polarization}g,h,i) corresponding to $H=4.24\;\mathrm{T}>H_c^\mathrm{1D}$ still features a longitudinal mode. As can be seen from the numerics the latter is entirely thermally activated. Similar thermal effects occur in gapped spin $S=1$ chains \cite{BeckerHonecker2017}. Interestingly, in Fig.~\ref{fig:polarization}i the thermally activated longitudinal mode is quite strong across the entire Brillouin zone, covering energies to about 0.5~meV, which is an order of magnitude larger than the temperature of the simulation. It only disappears when the temperature becomes small compared to the magnon gap, i.e., $k_\mathrm{B}T\lesssim g \mu_\mathrm{B}\mu_0 (H-H_c^\mathrm{1D})$, its intensity dropping across the entire zone. To our knowledge, such robustly thermally-activated longitudinal modes have not been observed for any quantum spin system to date.

\subsection{Interchain interactions}
The above discussion and a comparison with experiment (Fig.~\ref{fig:Cuts}) helps us evaluate the respective strengths and weaknesses of our two approaches to dealing with interchain coupling.
Chain-MF/RPA well reproduces the magnon energies, since it accounts for their dispersion transverse to the spin chains.
It does fail, though, in recovering longitudinal fluctuations at low temperatures, $k_BT\lesssim J_b$. This is because the one-dimensional model underlying chain-MF/RPA is gapped, while the actual quasi-2D system is not (Table.~\ref{tab:MagnetizationValues}). As a result, the reduction of magnetization due to a thermal activation of magnons is underestimated, and the resulting
magnetization is closer to saturation. Longitudinal fluctuations, which disappear altogether at full saturation, are then also underestimated.
At $k_BT> J_b$, the magnon activation gap becomes less important, the MF value of magnetization is closer to the correct one, and the calculation does a much better job with longitudinal excitations.

For the same reason our second approach is able to much better recover the intensity of the longitudinal mode even at low temperatures: the correct value of magnetization
is built into the method. Of course, being a purely one-dimensional calculation, it fails to take into account the magnon dispersion transverse to the spin chains. The predicted magnon energies are therefore slightly off.

\subsection{Scaling of critical fluctuations} \label{subsubsec:Scaling}

\begin{figure}
\includegraphics{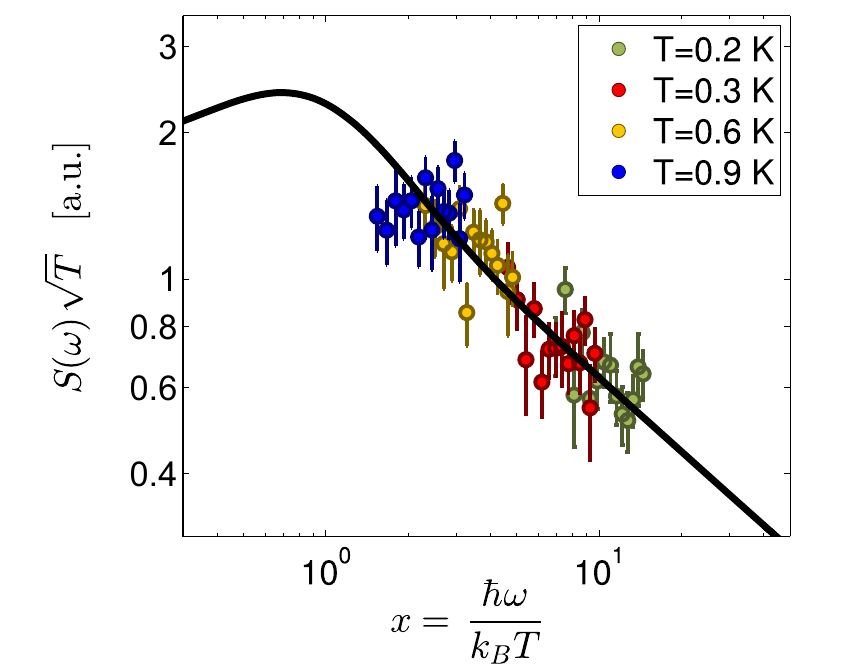}
\caption{\label{fig:Scaling} Scaling plot of the local dynamic structure factor. The data taken into account for this analysis is marked by the red dashed rectangles in Fig.~\ref{fig:Spectra}. The black line corresponds to the predicted universal scaling function.}
\end{figure}

For our initial quest of measuring critical scaling, the presence of longitudinal excitations constitutes a major problem since at this QCP only the transverse spin correlations become critical. We can only restrict the data range, focusing on energies and momenta where the longitudinal contribution is weak. This becomes progressively difficult at high temperatures, as longitudinal scattering broadens and intensifies. This restriction would be entirely absent in a polarized neutron study where longitudinal and transverse excitations can be separated. For a high polarization selectivity one would employ a three-axis spectrometer. Unfortunately, the point-by-point data collection in this technique, and a typical 20- to 40-fold intensity loss in the polarizer and analyzer, would realistically only allow measurements at one or a few select reciprocal space points. Polarized measurements over a broad momentum range and at several temperatures would take an unrealistic amount of beam time.

An even more obvious problem for us is interchain coupling, which spoils the one-dimensional nature of the QCP \cite{chitra_giam_qcp_universal_function_ll,giam_tsvelik_qcp_ladders_interplay_coupling,orignac_citro_giam_dimer_criticality}. The solution is to access a regime of high enough temperatures or frequencies at which the system will be dominated by the 1D QCP and not by 2D or 3D interactions. In practice, either the temperature or energy transfer should thus be high compared to the strength of one-dimensionality-breaking and anisotropic terms in the Hamiltonian \cite{Lake2005,Haelg2015_Ising,Haelg2015_CuDCl}. To avoid lattice effects, of course both temperatures and frequencies should stay small compared to the in-chain coupling and momentum transfers should be restricted to a regime where the magnon dispersion can be approximated as parabolic, as opposed to $\cos$-shaped \cite{chitra_giam_qcp_universal_function_ll,Lake2005}.
This approach is well tested and has been used in the context of Heisenberg spin chains \cite{Lake2005,Haelg2015_CuDCl},
ladders \cite{Povarov2015}, and Ising spin chains \cite{Haelg2015_Ising}.

Unfortunately, all these requirements combined leave us a quite restricted window of measurement at just a few experimental temperatures, as indicated by the red rectangles in Fig.~\ref{fig:Spectra}. Nevertheless this allows for a test of the quantum
critical scaling. For these temperatures and energy-momentum regions, we integrate our data in momentum to obtain the local dynamic structure factor
\begin{equation}
\mathcal{S}(\omega) = \int \mathcal{S}(\mathbf{q},\omega) \, \mathrm{d}\mathbf{q}.
\end{equation}

An attempted scaling plot for this quantity is shown in Fig.~\ref{fig:Scaling}. The temperature exponent is ${\gamma=-1/2}$ as predicted for the $z=2$, $d=1$ QCP \cite{sachdev_senthil_shankar_qcp_z2}. It is smoothly related to the exponent corresponding to the massless Tomonaga-Luttinger liquid which exists between the two saturation fields \cite{chitra_giam_qcp_universal_function_ll,giam_tsvelik_qcp_ladders_interplay_coupling} and corresponds to a theory of free particles. This can be understood on a physical basis due to the dilution of the excitations close to the QCP.
In other words, all the universal properties and scaling functions correspond to the properties of a one-dimensional gas of hard-core bosons or free spinless fermions.

For the scaling function of the local dynamic structure factor, the large- and small $x=\frac{\hbar \omega}{k_\mathrm{B}T}$ asymptotics were computed in Ref.~\onlinecite{sachdev_senthil_shankar_qcp_z2}. However, analytic expressions for the correlations were since derived \cite{korepin_slavnov_fredholm_string_scaling_function,zvoranev_cheianov_giam_fredholm_string}. Thus the scaling function can be directly obtained by numerically evaluating expressions obtained in Refs.~\onlinecite{korepin_slavnov_fredholm_string_scaling_function, sachdev_senthil_shankar_qcp_z2}. In part, such functions were published in Ref.~\onlinecite{BarthelArxiv2012Scalingfunctions}. In Appendix ~\ref{sec:ScalingFunctions} we use this approach to obtain the full scaling functions. The result is plotted as a solid line in Fig.~\ref{fig:Scaling} with an arbitrary overall scale factor.

The overlap between experimental data sets collected at different temperatures is of course too narrow to firmly claim as proof of scaling behavior. However, considering the severe intrinsic, instrumental and materials related limitations, the agreement between the theoretical scaling function and experiment is quite rewarding.

\section{Conclusion} \label{sec:Summary}

Using a combination of state-of-the-art high-resolution neutron spectroscopy and finite-temperature DMRG calculations we could
provide a detailed analysis of the excitations of a quantum spin chain near its quantum critical point at finite temperatures. The agreement between the experimental data and the calculations allowed to separate the various modes and thus to identify
properly the part of the signal to integrate to test the scaling function at the quantum critical point. The
theoretically-predicted finite temperature scaling of transverse spin correlations has been found to be in very good agreement
with the experimental data. In addition we observed a novel robust thermally activated longitudinal mode
persisting even to quite high temperatures.

\begin{acknowledgments}
We would like to thank D. Schmidiger and M. H\"alg for their involvement at the early stages of this project and B. Wehinger for providing quantum Monte Carlo simulations of the magnetization.
We thank P. Barmettler and C. Kollath for their involvement in the development of the DMRG code used in this study.
We acknowledge the sample environment team of the ISIS facility for their excellent support during the experiment.
This work is partially supported by the Swiss National Science Foundation under Division II.
\end{acknowledgments}

\appendix

\section{\label{sec:unpolarized_neutron}Weight of spectral contributions to the measured scattering intensity}
For unpolarized neutrons, the magnetic neutron scattering cross-section is proportional to the following weighted sum of spin structure factors \cite{Furrer2009}:
\begin{equation}
\frac{\mathrm{d}\sigma}{\mathrm{d}\Omega \mathrm{d}\omega} \propto \frac{k'}{k}F^2(\mathbf{Q}) e^{-2W(\mathbf{Q})} \sum_{\alpha, \beta} \left( \delta_{\alpha,\beta} -\frac{Q_\alpha Q_\beta}{Q^2}\right) \mathcal{S}^{\alpha \beta}(\mathbf{Q},\omega).
\end{equation}
Here $k',k$ denote the neutrons initial and final wave vector and this prefactor was factored out of the data. Also, our data has been corrected for the magnetic form factor $F(\mathbf{Q})$ of the Cu$^{2+}$ ion which is known and tabulated\cite{Brown2006}. Since all measurements were done at very low temperatures and low momentum transfer, the Debye-Waller factor $e^{-2W(\mathbf{Q})}$ which is due to atomic motion is merely a constant. Furthermore, choosing a basis such that the dynamic spin structure factor is diagonal the corrected scattering intensity is proportional to
\begin{equation}
\mathcal{I} \propto \sum_{\alpha } \left( 1 -\frac{Q_\alpha^2}{Q^2}\right) \mathcal{S}^{\alpha \alpha}(\mathbf{Q},\omega).
\end{equation}
For our Heisenberg spin chain system at saturation, it is only the external magnetic field along $z$ which breaks spin rotation symmetry. In the plain perpendicular to the external field we still have $\mathcal{S}^{xx}(\mathbf{Q},\omega) = \mathcal{S}^{yy}(\mathbf{Q},\omega)$. Therefore we can correct the measured data for the momentum dependent weight of the transverse contribution. Now, the corrected scattering intensity is proportional to the following weighted sum of correlators:
\begin{equation}
\mathcal{I} \propto \frac{1}{2}\left(\mathcal{S}^{xx}(\mathbf{Q},\omega)+ \mathcal{S}^{yy}(\mathbf{Q},\omega)\right) + \frac{Q^2 - Q_z^2}{Q^2+Q_z^2} \mathcal{S}^{zz}(\mathbf{Q},\omega).
\end{equation}
Thus the corrected neutron scattering intensity is {\it directly} proportional to the transverse dynamic spin structure factor and only the weight of the longitudinal contribution depends on wave-vector transfer. The volume of reciprocal space covered in our experiments is such that the momentum dependent weight of the longitudinal contribution is essentially unity in almost all of our data.

\section{\label{sec:BelowSaturation}Neutron scattering data at $H<H_0$}
\begin{figure}[h]
\includegraphics{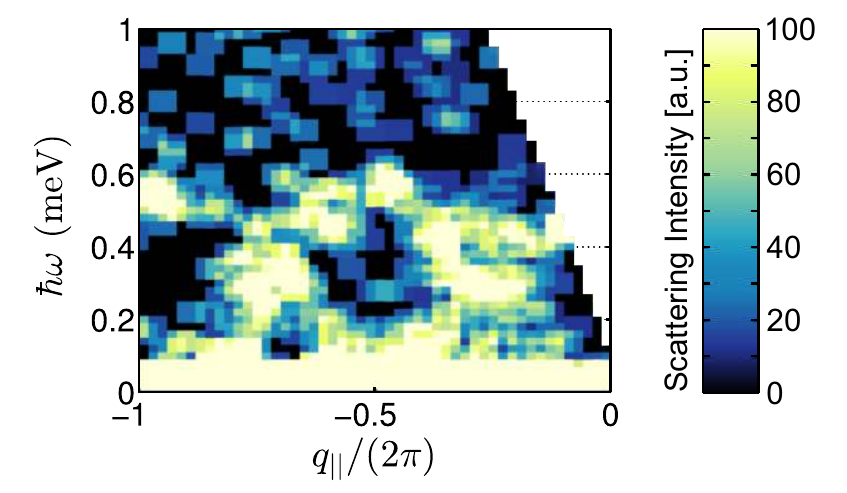}
\caption{\label{fig:4p07TeslaMmt} False-color plot of the neutron scattering intensity measured at $H = 4.07$~T at $T=0.2$~K.}
\end{figure}

Considering the data shown in Fig.~\ref{fig:GapAndDispersion}, it might appear, that the applied field of $H=4.47$~T is higher than the critical field. Then the actual critical field would be at $H=3.87$~T where the magnon gap extrapolates to zero. However, an additional short (low counting statistics) measurement at $H=4.07$~T shows that this is not the case. The obtained spectrum shown in Fig. \ref{fig:4p07TeslaMmt} clearly corresponds to a partially magnetized chain well below saturation. This conclusion is drawn from a comparison with Ref.~\onlinecite{Kohno2009} where such spectra have been numerically calculated for partially magnetized Heisenberg spin chains. 

\section{\label{sec:processing}Numerical simulations}

\subsection{\label{sec:about_tdmrg}T-DMRG }

The DMRG code is the one developed and used in Refs. \cite{coira_giam_kollath_spin_lattice_relax_nmr_low_t,coira_giam_kollath_dimerized_2bePublished} and based on a code initially developed by P. Barmettler.
The 1D simulations ran over $120$ sites with open boundary conditions on the spin-$\frac{1}{2}$ Heisenberg model (\ref{eq:1dheisenberg}) with coupling constant $J_\parallel$ and magnetic field $h^z$ spanning all the possible values $h^z_\mathrm{MF}(T)$ and $h^z_\mathrm{eff}(T)$ of Table \ref{tab:MagnetizationValues}.  For each different value, the T-DMRG converged to a thermal equilibrium distribution $e^{-\beta H/2}$ easily. Then, the dynamics was enabled by using the same procedure in real time after applying the observable $S^+_j$, $S^-_j$ or $S^z_j$ on the middle of the chain ($j=60$). The time evolution was restricted to $\chi=800$ renormalization dimension \cite{vidal_tebd_precursor_pure_state_dynamics_breakthrough} and the error then grows but remains controlled $|\epsilon_t|<10^{-8}$ up to time $t_\mathrm{max} = 30 J_\parallel^{-1}$. After each $\delta t = 0.05 J_\parallel^{-1}$ step, the dynamical averages and correlations over all sites were measured for temperatures listed in Table \ref{tab:MagnetizationValues}. We use overall the fourth Suzuki-Trotter decomposition \cite{schollwock_dmrg_long_notes} and the method described above is the scheme B in Ref.~\onlinecite{barthel_tdmrg_optimized_schemes}.  The chosen parameters allowed us to control the slightly growing error problem encountered in those algorithms and end with a better momentum and energy resolution than the experimental data.

\subsection{\label{sec:chi}Dynamical structure factor and retarded susceptibility}

The numerical T-DMRG simulations provide essentially exact values for the finite temperature spin correlation functions $\langle S^\alpha(r,t) S^\beta(0,0) \rangle$ with indices $\alpha \beta = \{+-,-+,zz \}$. Using the properties of the correlations and the symmetries of the Hamiltonian, we are now in the position to relate the commutators with the direct correlations
\begin{equation}
\left\langle \left[ S^z(r,t), S^z(0,0) \right] \right\rangle  =  2 i \, \mathrm{Im}\langle S^z(r,t) S^z(0,0) \rangle
\end{equation}
\begin{align}
\left\langle \left[ S^\pm(r,t), S^\mp(0,0) \right] \right\rangle &=  \langle S^\pm(r,t) S^\mp(0,0) \rangle \\
& \quad -    \langle S^\mp(r,t) S^\pm(0,0) \rangle^*
\end{align}
From these correlations, we can extract the retarded susceptibility
\begin{equation}
\chi^{\alpha\beta}_\mathrm{ret}(q,\omega)=  \int_{-\infty}^\infty   \mathrm{d}r  \, \mathrm{d}t  \,        - i \Theta(t) \left\langle \left[ S^\alpha(r,t), S^\beta(0,0) \right] \right\rangle e^{i(\omega t - q r)}
\label{eq:Susceptibility}
\end{equation}
where $\Theta$ is the Heaviside function. This retarded susceptibility corresponds to the linear response of the observable $S^\alpha({q},\omega)$ to a small
$\delta h$ perturbation of type $\delta h^\beta({q},\omega)
S^\beta({q},\omega)$.

The dynamical structure factor measured by the neutrons (\ref{eq:dynamic_struct_factor}) is related to the retarded susceptibility as follows
\begin{equation}
\mathcal{S}^{\alpha \beta}(\mathbf{q},\omega) = \frac{-2}{1-e^{-\beta \omega}} \Im{ \chi^{\alpha \beta}_{ret}(\mathbf{q},\omega)}.
\end{equation}

\subsection{\label{sec:filter}Finite size effects and applied filters}

The extracted correlations are bounded in space by the length of the system and in time by the entanglement growth of the dynamics. Thus any small numeric discontinuity at the border of our correlations ($\pm d_\mathrm{max}$ and $\pm t_\mathrm{max}$) would lead to artificial oscillating artifacts in susceptibility after Fourier transformation.

To avoid the finite size oscillations and the boundary effects, a filter mask has to be applied in space and time. For this the Gaussian filter
\begin{equation}
 M(x,t)=e^{-(A \, x/d_\mathrm{max})^2}e^{-(B \, t/t_\mathrm{max})^2}
\end{equation}
was used with $A=4$, $B=4$.

\begin{widetext}

\section{\label{sec:ScalingFunctions}Numerical computation of scaling functions}

The $z=2$, $d=1$ quantum critical point associated with the saturation of the magnetization can be described by a theory of nearly free excitations\cite{sachdev_senthil_shankar_qcp_z2}. In a bosonic language, close to the quantum critical point one is going to the limit of hard core (infinite local repulsion) bosons \cite{sachdev_senthil_shankar_qcp_z2}. Another transparent way to understand this limit is to perform a Jordan-Wigner transformation to map the excitations to fermions. In that case close to the QCP one is going to a limit of free spinless fermions \cite{chitra_giam_qcp_universal_function_ll}.  

The hard core boson limit was previously used to obtain the scaling functions for the equal time dynamic structure factor\cite{sachdev_senthil_shankar_qcp_z2}.  Here, we numerically evaluate the analytic expressions for the spin-spin correlation functions derived in Ref. \onlinecite{korepin_slavnov_fredholm_string_scaling_function} to obtain the dynamical scaling function. 

\subsection{Analytical expression for the correlation function}

The Hamiltonian considered by V. E. Korepin and N. A. Slavnov in Ref.~\onlinecite{korepin_slavnov_fredholm_string_scaling_function} reads (equations taken from this paper are tagged `KS' with the original equation number):
\begin{equation}
\mathcal{H}=\int_{-L/2}^{L/2} \mathrm{d}x \left(  \frac{\hbar^2}{2m}   \partial_x \Psi^\dagger    \partial_x \Psi      +   c \Psi^\dagger \Psi^\dagger \Psi \Psi        - h \Psi^\dagger \Psi  \right)
    \tag{KS 1.1}
\label{eq:KS-1p1}
\end{equation}
for a Bose field $\Psi(x)$ satisfying the usual bosonic commutation relations where only the case of hard core repulsion $(c\rightarrow \infty)$ is considered and $h$ denotes the chemical potential. In the following we set $\hbar=1$, $k_\mathrm{B}=1$ and $m=1/2$ to be consistent with Ref.~\onlinecite{korepin_slavnov_fredholm_string_scaling_function}. In the thermodynamic limit, at finite temperature, the following result is obtained:
\begin{equation}
    \langle \Psi(x_2,t_2) \Psi^\dagger(x_1,t_1) \rangle  =  e^{i h t_{21}} \left(  \frac{1}{2\pi} G(t_{12},x_{12})   +   \frac{\partial}{\partial \alpha} \right)  \det(1+V) \Bigr|_{(\alpha=0)}
    \tag{KS 4.8}
\label{eq:KS-4p8}
\end{equation}
\begin{equation}
    V(\lambda,\mu)=\frac{V_0(\lambda,\mu)}{\sqrt{1+e^{(\lambda^2-h)/T}}     \sqrt{1+e^{(\mu^2-h)/T}}  }
    \tag{KS 4.9}\label{eq-KS-4.9}
\end{equation}
\begin{equation}
\begin{aligned}
    V_0(\lambda,\mu)=  \left[  \frac{  E(\lambda | t_{12},x_{12})  -  E(\mu | t_{12},x_{12})   }{ \pi^2  (\lambda-\mu)}      -        \frac{\alpha}{2\pi^3} E(\lambda | t_{12},x_{12})   E(\mu | t_{12},x_{12})      \right]     \\
\cdot  \exp \left(\frac{i}{2}t_{21}(\lambda^2+\mu^2) - \frac{i}{2} x_{21}(\lambda+\mu)\right)
\end{aligned}
    \tag{KS 3.12}\label{eq-KS-3.12}
\end{equation}
\begin{equation}
    E(\lambda|t,x)=\dashint_{-\infty}^\infty     \frac{e^{i t \mu^2 - i x \mu}}{\mu-\lambda} \,   \mathrm{d}\mu
    \tag{KS 3.6}\label{eq-KS-3.6}
\end{equation}
\begin{equation}
    G(t,x)=\int_{-\infty}^\infty e^{i t \mu^2 - i x \mu} \,   \mathrm{d}\mu
    \tag{KS 3.9}\label{eq-KS-3.9}
\end{equation}
Here, $t_2-t_1=t_{21}>0$, $t_{12}=-t_{21}$ and $x_2-x_1=x_{21}>0$, $x_{12}=-x_{21}$. The dashed integration sign denotes the Cauchy principal value integral. Note that in Ref. \onlinecite{korepin_slavnov_fredholm_string_scaling_function} there is a minus sign error in equation \ref{eq-KS-3.6}. This is evident by comparison to Eq. (4.6) in the same paper and by comparison to Eq. (6.2) in Sec. XIII of Ref. \onlinecite{KorepinBogoliubovIzergin1993}.

\subsection{Numerical evaluation of the correlation function}
In order to numerically evaluate the correlation function \ref{eq:KS-4p8}, we need to discretized the operator $V$ into a matrix, numerically obtain this matrix and deal with the derivative with respect to $\alpha$.

\subsubsection{Derivative of the determinant}
Note that $V$ is linear in $\alpha$. We define $V=V_1+\alpha V_2$.  Using $ \frac{\mathrm{d} \det(A)}{\mathrm{d}\alpha} = \det(A) \,  \mathrm{tr} \left( A^{-1} \frac{\mathrm{d}A}{\mathrm{d}\alpha} \right)  $ we find
\begin{equation}
\begin{aligned}
     \frac{\partial}{\partial \alpha}  \det(1+V) \Bigr|_{(\alpha=0)} & =
\frac{\partial}{\partial \alpha}  \det(1+V_1 + \alpha V_2) \Bigr|_{(\alpha=0)}  =  \det(1+V_1) \,  \mathrm{tr} \left(   (1+V_1)^{-1} V_2 \right).
\end{aligned}
\end{equation}

\subsubsection{Analytic expressions for $V$}
When numerically evaluating $V(\lambda,\mu)$ using \ref{eq-KS-4.9}, \ref{eq-KS-3.12}, \ref{eq-KS-3.6} and \ref{eq-KS-3.9}. There are two difficulties. 

\begin{enumerate}
  \item The integrals in $E(\lambda|t,x)$ and $G(t,x)$ must be brought into a form that is amendable to rapid numerical evaluation:
\begin{equation}
    G(t,x)=\int_{-\infty}^\infty e^{i t \mu^2 - i x \mu} \,   \mathrm{d}\mu =     \sqrt{\frac{\pi}{2 |t|}}  \left( 1+ \mathrm{sign}(t) i \right)      \, e^{-\frac{i x^2}{4 t}}  \qquad \text{for $t\neq0$}
\end{equation}
\begin{equation}
  E(\lambda|t,x)  =\dashint_{-\infty}^\infty     \frac{e^{i t \mu^2 - i x \mu}}{\mu-\lambda} \,   \mathrm{d}\mu                
=  -  i \pi  \,\mathrm{erf}\left(\frac{(1+\mathrm{sign}(t) i)}{2\sqrt{2}} \frac{(x - 2 \lambda  t)}{\sqrt{|t|}}\right)  e^{i \lambda ^2 t-i  \lambda x} \qquad \text{for $t\neq0$}
\label{eq:E}
\end{equation}
where $\mathrm{erf}(z)$ is the error function, which can be efficiently evaluated numerically.

\item Care must be taken when evaluating the diagonal entries ($\lambda = \mu$). In Eq. \ref{eq-KS-3.12}, one should take the limit of the first part carefully:
\begin{align}
\lim_{\lambda \to \mu}   \left(    \frac{  E(\lambda | t_{12},x_{12})  -  E(\mu | t_{12},x_{12})   }{ \pi^2  (\lambda-\mu)}    \right)  &=  \frac{1}{\pi^2} \,  \frac{\partial}{\partial \lambda}  E(\lambda | t_{12},x_{12})    \Bigr|_{(\lambda = \mu)} \notag \\ 
& = \frac{e^{i \mu ^2 t-i \mu  x}}{\pi^2}  
 \left(4 i \sqrt{\pi } \gamma  t \, e^{-\gamma ^2 (x-2 \mu  t)^2}-  \pi  (x-2 \mu  t)  \text{erf}(\gamma  (x-2 \mu  t))\right),
\end{align}
where $\gamma = \frac{(1+\mathrm{sign}(t) i)}{2\sqrt{2} \sqrt{|t|}}$.
\end{enumerate}

\subsubsection{Discretization of $V$}
The analytic expressions correspond to the thermodynamic limit, i.e., infinite system size and infinite number of particles. For the numerical evaluation, the momenta were discretized to an equally spaced grid
\begin{equation}
     \lambda_{j}, \mu_{j}=\frac{2 \pi}{L} \left(  j - \frac{N+1}{2}  \right), \qquad j=1,\ldots,N.
\end{equation}
The operator $V(\lambda,\mu)$ then becomes a matrix $\tilde{V}_{i,j}= \frac{2 \pi}{L}    V(\lambda_{i},\mu_{j}) $.

\subsection{Dynamic structure factor}
Having evaluated the full space- and time-dependent finite-temperature correlation function, the dynamic structure factor is obtained by Fourier transformation.
\begin{equation}
    g(x,t) := \langle \Psi(x_2,t_2) \Psi^\dagger(x_1,t_1) \rangle
\end{equation}
\begin{equation}
    \mathcal{S}_{+-}(k,\omega) = \int e^{- i k x + i \omega t}\,  g(x,t)    \,\, \mathrm{d}x \, \mathrm{d}t
, \qquad           \mathcal{S}_{-+}(k,\omega) =   e^{\hbar \omega / k_\mathrm{B}T}\, \mathcal{S}_{+-}(-k,-\omega)
    \label{eq:StructureFactor},
\end{equation}
where $x=x_2-x_1$ and $t=t_2-t_1$.

\subsection{Numerical results}
We have numerically evaluated the real-space and real-time correlation functions at different temperatures and chemical potentials. The obtained dynamic structure factors $\mathcal{S}^{+-}(k,\omega)$ are shown in Fig.~\ref{fig:AllHcbSpectra} as false color plots\footnote{The computed real space and time correlation functions were multiplied by a broad Gaussian $\exp[-(x^2+t^2)/2\sigma^2]$ with $\sigma=20$. After Fourier transformation, this corresponds to a convolution of the dynamic structure factor with a narrow Gaussian of width $\tilde{\sigma}=0.05$ }. As expected, at negative chemical potentials (no particles at $T=0$) we observe a gapped parabolic dispersion. At positive chemical potential i.e. non-zero boson density in the ground state we see linearly dispersive excitations corresponding to the Tomogana-Luttinger liquid regime.

The local dynamic structure factor $\mathcal{S}(\omega)$ is obtained by Fourier transformation of the local $(x=0)$ real-time correlation function. This is done in Fig.~\ref{fig:HardCoreBosons_LDSF} for different points in the $(T,h)$-phase diagram. Here, making use of relation (\ref{eq:StructureFactor}) we compute the full local dynamic structure factor $\mathcal{S}=\frac{1}{2}(\mathcal{S}_{+-}+\mathcal{S}_{-+})$. The obtained curves for $\mathcal{S}(\omega)$ are shown in Fig.~\ref{fig:HardCoreBosons_LDSF}b. If plotted in scaled variables (Fig.~\ref{fig:HardCoreBosons_LDSF}c) all curves for a fixed ratio of the chemical potential to temperature $r=h/T$ collapse onto a single line as predicted in Ref. \onlinecite{sachdev_senthil_shankar_qcp_z2}. It is this $r=0$ scaling function that is used in the main text for comparison with our experimental data.

\begin{figure*}[h]
\includegraphics{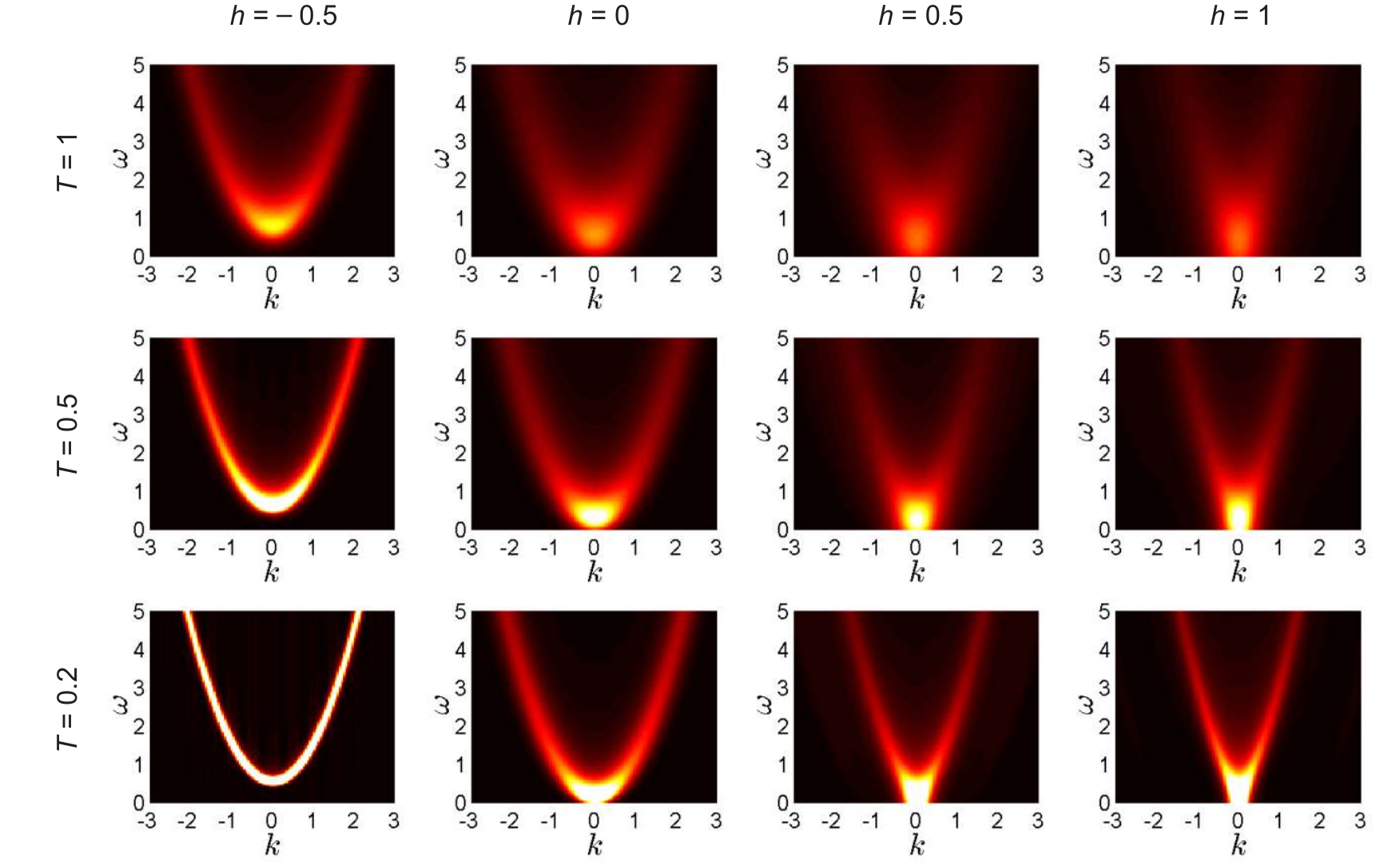}
\caption{\label{fig:AllHcbSpectra} False color plots of the dynamic structure factor $\mathcal{S}^{+-}(k,\omega)$ for impenetrable bosons in one dimensions at different chemical potentials $h=-0.5,0,0.5,1$, and temperatures $T=0.2,0.5,1$. }
\end{figure*}

\begin{figure*}[h]
\includegraphics{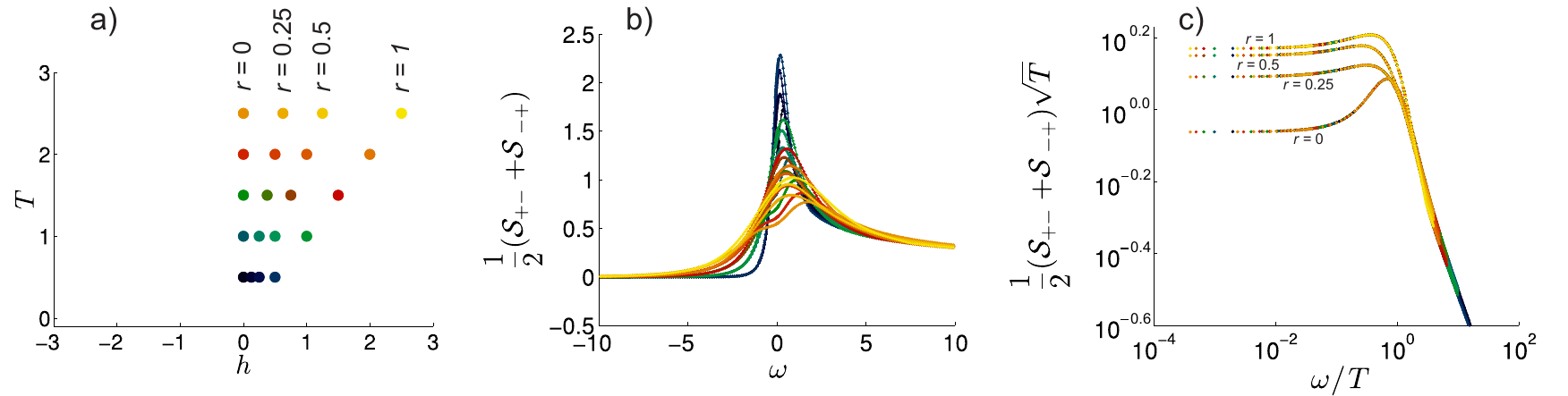}
\caption{\label{fig:HardCoreBosons_LDSF} Local dynamic structure factor calculated for different points in the $(T,h)$-phase diagram (a) and plotted in real (b) and in scaled variables (c). As predicted in Ref.~\onlinecite{sachdev_senthil_shankar_qcp_z2}, all curves for a fixed ratio of the chemical potential to temperature $r=h/T$ collapse onto a single curve if plotted in appropriately scaled variables.}
\end{figure*}

\end{widetext}

\bibliography{references,noam}

\end{document}